\documentclass[sigconf]{acmart}

\usepackage{booktabs} 
\usepackage{multirow}
\usepackage{adjustbox}
\usepackage{subcaption}
\usepackage{todonotes}
\usepackage{framed}
\usepackage[flushleft]{threeparttable} 
\setcopyright{none}

\acmConference[ESEM]{International Symposium on Empirical Software Engineering and Measurement}{October 2018}{Oulu, Finland}

\acmPrice{0.00}

\newcommand{\MYCOMMENT}[1]{}

\usepackage{xspace}

\newcommand{\ie}{\textit{i.e.,}\xspace}
\newcommand{\eg}{\textit{e.g.,}\xspace}

\newcommand{\etal}{\textit{et al.}\xspace}

\newcommand{\tdd}{TDD\xspace}
\newcommand{\yw}{YW\xspace}

\newcommand{\mra}{MRA\xspace}
\newcommand{\bsk}{BSK\xspace}
\newcommand{\ssh}{SSH\xspace}
\newcommand{\gol}{GOL\xspace}

\newcommand{\qlty}{QLTY\xspace}
\newcommand{\pro}{PROD\xspace}
\newcommand{\test}{TEST\xspace}

\begin{document}
\title{A Longitudinal Cohort Study on the Retainment of Test-Driven Development}

\author{Davide Fucci}
\affiliation{%
  \institution{University of Hamburg}
  \city{Hamburg}
  \country{Germany}
}
\email{fucci@informatik.uni-hamburg.de}

\author{Simone Romano}
\affiliation{%
  \institution{University of Basilicata}
  \city{Potenza}
  \country{Italy}
}
\email{simone.romano@unibas.it}

\author{Maria Teresa Baldassarre}
\affiliation{%
  \institution{University of Bari}
  \city{Bari}
  \country{Italy}
}
\email{mariateresa.baldassarre@uniba.it}

\author{Danilo Caivano}
\affiliation{%
  \institution{University of Bari}
  \city{Bari}
  \country{Italy}
}
\email{danilo.caivano@uniba.it}

\author{Giuseppe Scanniello}
\affiliation{%
  \institution{University of Basilicata}
  \city{Potenza}
  \country{Italy}
}
\email{giuseppe.scanniello@unibas.it}

\author{Burak Thuran}
\affiliation{%
  \institution{Brunel University}
  \city{London}
  \country{UK}
}
\email{burak.turhan@brunel.ac.uk}

\author{Natalia Juristo}
\affiliation{%
  \institution{Universidad Politecnica de Madrid}
  \city{Madrid}
  \country{Spain}
}
\email{natalia@fi.upm.es}

\renewcommand{\shortauthors}{Fucci \etal}

\begin{abstract}
\textit{Background:} Test-Driven Development (\tdd) is an agile software development practice, which is claimed to boost both external quality of software products and developers' productivity. 

\noindent
\textit{Aims:} We want to study: \textit{(i)} the \tdd effects on the external quality of software products as well as the developers' productivity; and~\textit{(ii)}~the retainment of \tdd over a period of five months. 


\noindent
\textit{Method:} We conducted a (quantitative) longitudinal cohort study with 30 third-year undergraduate students in Computer Science at the University of Bari in Italy. 

\noindent
\textit{Results:} The use of \tdd has a statistically significant effect neither on the external quality of software products nor on the developers' productivity. 
However, we observed that participants using \tdd produced significantly more tests than those applying a non-\tdd development process, and that the retainment of \tdd is particularly noticeable in the amount of tests written.


\noindent
\textit{Conclusions:} Our results should encourage software companies to adopt \tdd because who practices \tdd tends to write more tests---having more tests can come in handy when testing software systems or localizing faults---and it seems that novice developers retain \tdd.


\end{abstract}

%
%



\maketitle

\section{Introduction}\label{sec:introduction}
Test-Driven Development (TDD) is a software development technique which leverages unit tests to incrementally deliver small pieces of functionality.
Peculiar to TDD is the order in which tests and production code are written---the former are specified first and only in the case of a failure the developer is allowed to write production code to make them pass~\cite{Bec03}.
An important role in this process is played by refactoring which allows changing the code internal representation (\eg an algorithm) while preserving its external behavior due to the safety net provided by the test~suite~\cite{Ast03}.

TDD promised to support the delivery of high-quality products, both from a functional (\eg fewer bugs) and technical perspective (\eg ``cleaner'' code), while improving developers' productivity~\cite{Bec03}.
Consequently, industry has taken an interest in adopting this technique~\cite{CSP11} and academia has dedicated large effort to gather empirical evidence to support or disprove its claimed effects. 
The results, gathered and combined in secondary studies~\cite{MMP14,RM13,TLD06}, are conflicting and limited conclusions can be drawn. 
The primary studies, such as controlled experiments and case studies, are often cross-sectional~\cite{CCS02} and only capture a ``snapshot'' of the phenomena at a given time. 
However, despite being recommended in the literature~\cite{SMT10, FTJ15, MH07}, only few investigations~\cite{Lat14, SWM07} take a longitudinal perspective on the study of TDD. 
In one of these studies, Latorre~\cite{Lat14} followed professional developers of different levels of seniority (but all with no experience in TDD) working on a project for a month while learning to apply the technique.
The author studied how the conformance to the TDD process and the participants' productivity evolved during the investigation.
The focus was to evaluate how the different subjects' learning curves affect their performance (\eg in terms of code quality).
A long-term case study at IBM by Sanchez \etal~\cite{SWM07} aimed at understanding whether TDD improves over the process previously adopted in the company.
The observation focused on a team and its sustained use of TDD for a period of five years.
However, the investigation was carried out retrospectively---\ie using existing data gathered during such period but also before TDD was introduced.  
Conversely, we present a longitudinal cohort study~\cite{CCS02} involving two separate observations of the same variables (\ie functional quality, productivity, and number of tests written), obtained from the same participants (\ie 30 novice developers), five months apart. 
Our cohort is composed of software developers of homogeneous experience who attended the same training regarding Agile software development principles, including TDD. 
Our goal is to understand how well TDD can be applied after the passage of time, giving an indication of its~retention.

Thus, the main research question driving our study is:
\begin{framed}
\noindent
To what extent can novice software developers \textit{retain} TDD and its effects (if any) over a period of five months? 
\end{framed}
To establish a baseline, we compared the treatment of interest (\ie TDD) with the non-TDD development process (\eg iterative test-last, big-bang testing, or no testing at all) that subjects would normally follow. 
We refer to the latter as \textit{Your Way} development (\ie~YW).

This paper makes the following main contributions:
\begin{itemize}
    \item an evidence-based discussion of TDD retainment and its implication for research and practice;
    \item a longitudinal design methodology that can be applied to other software development processes to distinguish between short-term and long-term phenomena;
    \item a laboratory package\footnote{\href{http://www2.unibas.it/sromano/downloads/LabPackageUniba.zip}{www2.unibas.it/sromano/downloads/LabPackageUniba.zip}} to foster further replication of the presented longitudinal cohort study.
\end{itemize}

\noindent
\textbf{Paper organization.} In Section~\ref{sec:background}, we present background information and related work. In Section~\ref{sec:study}, we describe the design of our longitudinal cohort  study, while the results are presented and discussed in Section~\ref{sec:results} and Section~\ref{sec:discussion}, respectively. Final remarks conclude the paper in Section~\ref{sec:conclusion}.
\section{Background}\label{sec:background}
In this section, we summarize the available evidence supporting (or refuting) the effects of TDD on external quality (or functional \eg number of defects) and developers' productivity.
We  also summarize research work on existing longitudinal studies in the context of Software Engineering (SE).

\subsection{Types of longitudinal studies in SE}
Longitudinal research in SE is not so common and appears to be mostly associated with the case study methodology. 
According to Yin~\cite{Yin09}, in a longitudinal case study data collection happens over an extended period with the goal of investigating \textit{``how certain conditions change over time''}~\cite{Yin09}.
This is the case when the phenomenon under investigation is a process bounded to its context. 
Therefore, similarly to ethnography~\cite{SDD10}, longitudinal case studies require the researchers to be co-located with the case in which the phenomena takes place.
For example, in the longitudinal case study reported by McLeod \etal~\cite{MMD11}, researchers spent several hundreds of hours at the case company site. 
They attended meetings, observed and interviewed stakeholders within a period of two years to characterize software development as an emergent process. 
In a similar fashion, Salo and Abrahamson~\cite{SA05} followed how Software Process Improvement (SPI) techniques were introduced in the workflow of five Agile projects.
Their investigation lasted for 18 months during which the researchers constantly recorded the output of retrospective meetings, interviews with the developers, as well as the metrics collected from the SPI tool in use at the company.

Longitudinal studies are useful when the observations cover an interesting event---\eg the introduction of a new practice within a company.
Therefore, the researcher is interested in observing the impact of such change while it unfolds.
This scenario is similar to interrupted time series in quasi-experimental designs~\cite{CCS02}.
For example, Li \etal~\cite{LMD10} studies the changes brought by replacing a waterfall-like approach with Scrum in a small software company.
The authors followed the development of a project for more than three years---17 months using waterfall and 20 using Scrum.
This approach allowed for a \textit{before-after} comparison of defects density and productivity. 
The longtime span was necessary to avoid a biased comparison between the established process and an immature one.
A similar approach is reported in Vanhanen \etal~\cite{VLM07} in which the impact of introducing pair programming was assessed over a period of two years with data collected through survey with the developers. 

Other examples of longitudinal studies in SE cover a long period of time in retrospect---\eg by analyzing archival data.
Harter \etal~\cite{HKS12} analyzed the type of defects identified over time by the progressive introduction of SPI techniques in a firm and its subsequent CMMI improvements over a period of 20 years.

In the health science and medicine, longitudinal studies are sometimes realized in the form of cohort studies. 
A cohort is a sample of subjects (\eg who undergo a treatment) sharing a specific characteristic of interest (\eg age).
The cohort is tested in several occasions over time to, for example, check for a drug side-effect before releasing it to the market~\cite{CCS02}.

\begin{figure*}[t]
	\centering
  	\includegraphics[width=\linewidth]{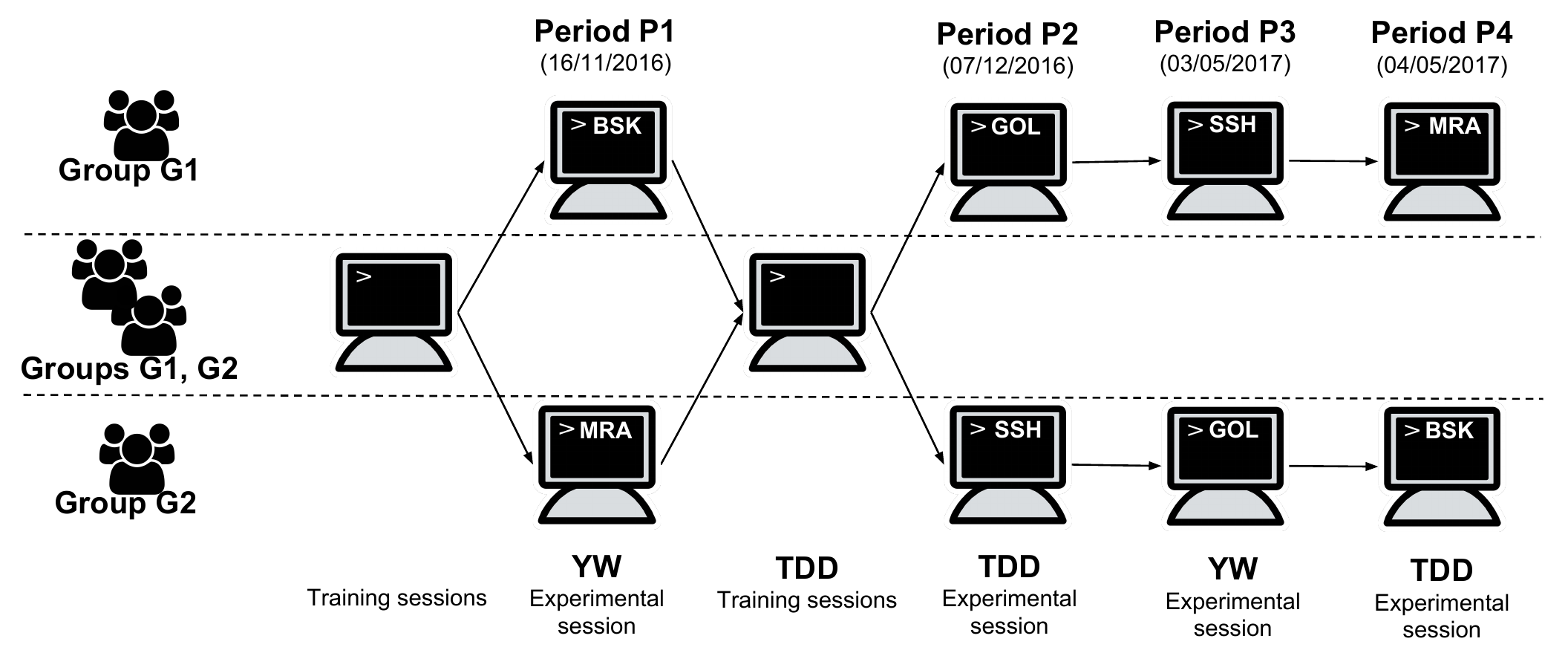}
    \caption{Summary of the study.}\label{fig:study}
\end{figure*}

\subsection{Effects of TDD}
The effects of TDD on several outcomes, including the ones of interest for this study---\ie functional quality and productivity---is the topic of several empirical studies, summarized in Systematic Reviews (SR) and Meta-Analysis (MA).
Turhan \etal~\cite{TLD06} SR includes 32 primary studies in which TDD was investigated in different settings. 
Although the results show a positive effect on quality, the ones regarding productivity are inconclusive.
Rafique and Misic~\cite{RM13} conducted an MA covering 25 primary studies published between 2000 and 2011. 
When considering participants from academia, TDD seems to improve quality to the loss of productivity. 
Finally, Munir \etal~\cite{MMP14} SR classifies the primary studies according to relevance and rigor dimensions.
The results show, for both student and professional developers, that TDD increases quality but not productivity. 
The authors recommend increasing relevance by carrying out long-term studies.
One example of such investigation is presented in Marchenko \etal~\cite{MAI09} which reports a three-year-long case study about the use of TDD at Nokia-Siemens Network. 
The authors observed and interviewed eight participants (one Scrum master, one Product owner, and six developers) and extracted themes from the data. 
The participants perceived TDD as an important driver towards the improvement of their code quality, both from a structural and functional perspective. 
Moreover, the team confidence with the code base improved, which is associated with improved productivity~\cite{Bec03}.
The examined team reported that TDD was not suitable for bug fixing, especially for bugs that are difficult to reproduce (\eg needing a specific environment setup) or for quick ``hacks'' due to the testing overhead.
The authors also report some concerns regarding the lack of a solid architecture when applying TDD.

Latorre~\cite{Lat14} studied the capability of 30 professional software developers (junior, intermediate, and experts) to develop a real-world system using TDD. 
The study targeted the \textit{learnability} of TDD, as the participants did not know the technique beforehand.
The longitudinal one-month study started after giving the developers, proficient in Java and unit testing, a tutorial about TDD.
After only a short practice session, the participants were able to correctly apply TDD (\eg following the prescribed steps).
Although they correctly followed the process between 80\% and 90\% of the time, their ability to initially apply TDD depended on experience---while seniors needed only few iterations, intermediates and juniors needed more time.
Experience had an impact on productivity---only the experts were able to be as productive as they were when applying a traditional development methodology (measured during the initial development of the system).
Refactoring and design decision hindered the productivity of intermediates and junior participants.
Finally, regarding functional quality, all the participants in the study delivered a correct version of the system regardless of their seniority~level. 


\section{Longitudinal Cohort Study}\label{sec:study} 
In this section, we describe the planning of our longitudinal cohort study. We summarize the most important steps of this study in Figure~\ref{fig:study}. In particular, the participants in the study (groups G1 and G2) first took part in training sessions (and accomplished homework) where they practiced unit testing, iterative test-last development, and big-bang testing. Then, the participants in the groups G1 and G2 were asked to perform two implementation tasks on two different experimental objects (\ie \bsk\footnote{\bsk (Bowling ScoreKeeper) is an API for calculating the score of a bowling game.} and MRA,\footnote{\mra (Mars Rover API) is an API for moving a rover on a planet.}) in the same period P1---a period is the time during which a treatment is applied~\cite{Vegas:2016}. In P1, the participants could apply only \yw because they were not aware of \tdd yet. Between the periods P1 and P2, all the participants (G1 and G2) practiced \tdd during training sessions (and homework). In the second period (\ie P2), we asked the participants in G1 and G2 to perform other two tasks, GOL\footnote{\gol (Game Of Life) is an API for Conway's game of life.} and SSH\footnote{\ssh (SpreadSHeet) is an API for a spreadsheet.} respectively. 
The applied treatment was \tdd for both groups. 
After five months, we asked the same participants in group G1 to implement a new task---i.e., \ssh---during P3 using the \yw approach. 
In the same period (\ie P3), the participants of G2  implemented the \gol task using \yw. 
While, during P4, the participants of G1 and G2 were asked to apply \tdd on \mra and \bsk, respectively. 
We considered P3 and P4 to study the effects of \tdd and its retainment. 
We introduced the last periods, P3 and P4, several months apart from the first two to assess whether the initial knowledge of \tdd is retained. 

The planning of our cohort study is reported according to the template by Jedlitschka \etal~\cite{Jedlitschka2008}. When planning and conducting our study, we followed the guidelines by Juristo and Moreno~\cite{Juristo:2001} and Wohlin \etal~\cite{Wohlin:2012}.

\subsection{Research Questions}
We aimed at investigating the following Research Questions (RQs):

\begin{description}
\item[RQ1.] To what extent do novice software developers retain \tdd and how does this affect their performance?
\item[RQ2.] Are there differences between \tdd and \yw in the external quality of the implemented solutions, developers' productivity, and number of tests written?  
\end{description}

We defined RQ1 to study whether \tdd retainment affects the application of \yw and whether there are deteriorations (or improvements) in the application of \tdd over five months. 
The considered constructs are, external quality of the implemented solutions, developers' productivity, and the number of tests developers wrote.     

Finally, RQ2 aimed at understanding whether the claim that \tdd increases both external quality of the software products and developers' productivity is well-founded as well as whether \tdd leads developers to write more tests. 

\subsection{Experimental Units}
The participants were third-year undergraduate students in Computer Science.
They were sampled by convenience among the students taking the \textit{Integration and Testing} course at the University of Bari in Italy.
The course covered the following topics, software quality, unit testing, integration testing, SOLID principles, refactoring, iterative test-last development, big-bang testing, and \tdd.
The course included frontal lectures, laboratory sessions, and homework.
During the laboratory sessions, the students improved their knowledge about how to develop unit tests in Java by using the Eclipse IDE and JUnit, and the refactoring functionality available in Eclipse. 
During laboratory sessions and by developing homework, the students practiced unit testing, iterative test-last development, big-bang testing, and \tdd.

Participation in the cohort study was voluntary.
We informed the students that any gathered data would be treated anonymously and used for research purposes only. We also informed them that their performance in the study would not affect their final mark for the Integration and Testing course.
To encourage the participation, we rewarded who accepted to take part in the study with a bonus in their final mark.
Among the students taking the Integration and Testing course, 30 accepted to participate. 

The participants had passed the exams for the courses of Procedural Programming, Object Oriented Programming, Software Engineering, and Databases.
During these courses, all participants had acquired programming experience in C and Java. 
Between the first two periods and the last two, the participants followed the same university curricula courses in which \tdd was not used.
However, we did not control whether, within such period, the participants practices \tdd outside the academic scope (\eg in personal projects). 

\subsection{Experimental Materials}
The experimental objects were four code katas (\ie programming exercises used to practice a technique or a programming language).
\begin{itemize}
\item \textbf{\bsk.} It is an API for calculating the score of a bowling game. This API allows adding a frame to a game, as well as bonus throws; computing the score of a frame; identifying when a frame is spare or strike; and computing the score of a~game.    
\item \textbf{\mra.} It is an API for moving a rover on a planet, which is represented as a grid. The cells of this grid can contain obstacles that the rover cannot pass through. \mra allows the initialization of a planet (\ie defining the grid with the obstacles) and moving the rover on the planet by parsing a list of commands (\ie turning left/right and moving forward/backward). 
\item \textbf{\ssh.} It is an API for a spreadsheet. \ssh allows evaluating the content of a cell and thus returning the result of this evaluation. Cells can contain integers, strings, references, and formulas (\eg concatenation of strings or integer addition).  
\item \textbf{\gol.} It is an API for Conway's game of life. This game takes place on a square grid of cells. Each cell can assume two states: alive or dead. At each step, the current state of the grid is used to determine the next state. \gol allows initializing the grid and determining the next state of each cell (it depends on the current state of the cell and of its neighbors) and then of the grid. 
\end{itemize}
The implementation of the aforementioned APIs did not require any graphical user interfaces. 

Each experimental object was composed of several user stories\footnote{A user story is a description of a feature to be implemented from the perspective of the end user.} to be implemented, as well as a template project (of Eclipse) that contained a stub of the expected API signature and an example JUnit class test. To verify that the user stories were correctly implemented, each experimental object was accompanied by acceptance test suites---an acceptance test suite for each user story. It is worth mentioning that the acceptance test suites were not provided to the participants. That is, these suites were only exploited to quantify the quality of the solutions implemented by the participants and their productivity (see Section~\ref{sec:var}).

The use of code katas in empirical studies on \tdd is quite common (\eg~\cite{Fucci:2016,Fucci:2017,Erdogmus:2005}). For \bsk and \mra, we exploited the materials used in the experiment by Fucci \etal~\cite{Fucci:2016}. As for \ssh and \gol, we created the experimental materials (\eg template projects). 

\subsection{Tasks}
Each task was coupled to an experimental object (\ie four tasks, one for each experimental object). A task consisted of implementing a solution for an experimental object (\eg \bsk). To this end, we provided the participants with: \textit{(i)} the user stories to be implemented for the considered experimental object; and \textit{(ii)} the template project.  Thus, the participants had to use the template project when implementing the user stories for that experimental object. 

\begin{table*}[t]
\centering
\caption{Design Summary.}
\label{tab:design}
\begin{tabular}{llcccc}
\toprule
& & \multicolumn{4}{c}{\textbf{Period}} \\
                   &  &  \textbf{P1 (16/11/2016)} & \textbf{P2 (07/12/2016) }& \textbf{P3 (03/05/2017)} &\textbf{ P4 (04/05/2017)} \\ \midrule
\multirow{2}{*}{Group} & G1 & YW,  BSK      & \tdd, GOL      & YW, SSH       & \tdd, MRA      \\
& G2 & YW,  MRA      & \tdd, SSH      & YW, GOL       & \tdd, BSK      \\ \bottomrule
\end{tabular}
\end{table*}

\subsection{Hypotheses and Variables}\label{sec:var}
 
The participants were asked to carry out each task by using either \tdd or the approach they preferred (\ie \yw)---of course, in this latter case, they could not use \tdd.  
Therefore, one of the independent variable (also known as factor) is \textbf{Technique}. It is a nominal variable assuming two values, \textit{\tdd} and \textit{\yw}.
Since our study is longitudinal---\ie we collected data over time--- we took into account another independent variable, which represents the period during which each treatment (\ie \tdd or \yw) was applied. We named this variable \textbf{Period}. It is a nominal variable and assumes the following values, \textit{P1}, \textit{P2}, \textit{P3}, and \textit{P4}. 
We also considered the independent variable \textbf{Group} representing the two groups of participants. It is a nominal variable assuming two values: \textit{G1} and \textit{G2}. 


The dependent variables considered in our study are, \textbf{\qlty}, \textbf{\pro}, and \textbf{\test}. 
We choose these dependent variables because they have been previously used in other empirical studies on \tdd (\eg~\cite{Erdogmus:2005,Fucci:2016,Fucci:2017,Tosun:2017}). The variable \qlty quantifies the external quality of the solution a participant implemented. This variable is defined as follows (\eg~\cite{Fucci:2016}):
\begin{equation}\label{eq:1}
\qlty=\frac{\sum_{i=1}^{\#TUS} \qlty_i}{\#TUS} * 100
\end{equation}
where \#TUS is the number of user stories a participant tackled, while QLTY\textsubscript{i} is the external quality of $i$-th user story. To understand if a user story was tackled or not, we checked the asserts in the corresponding acceptance test suite. Namely, if at least one assert in the test suite (for that story) passed, then the story was tackled. \#TUS is formally defined as:
\begin{equation}\label{eq:2}
\#TUS=\sum_{i=1}^n \bigg \{ \begin{array}{rl}
1 & \#ASSERT_i(PASS)>0 \\
0 & \textstyle otherwise \\
\end{array}
\end{equation}
On the other hand, the quality of the $i$-th user story (\ie QLTY\textsubscript{i}) is equal to the number of asserts passed for the $i$-th story with respect to the total number of asserts for the same story. More formally: 
\begin{equation}\label{eq:3}
QLTY_i=\frac{\#ASSERT_i(PASS)}{\#ASSERT_i(ALL)}
\end{equation}
Given the Formulas \ref{eq:1},\ref{eq:2}, and \ref{eq:3}, \qlty assumes values between 0 and 100, where a value close to 0 means that the quality of the solution is low, while a value close to 1 indicates high quality of the solution. 

The variable \pro estimates the productivity of a participant. It is computed as follows (\ie \cite{Tosun:2017}): 
\begin{equation}\label{eq:4}
\pro=\frac{\#ASSERT(PASS)}{\#ASSERT(ALL)}*100
\end{equation}
where \#ASSERT(PASS) is the number of asserts passed, by considering all the acceptance test suites, with respect to the total number of the asserts in the acceptance test suites. \pro assumes values between 0 and 100. A value close to 0 indicates low productivity, while a value close to 1 means high productivity.

The variable \test quantifies the number of unit tests a participant wrote. It is defined as the number of asserts in the test suite written by a participant when tackling a task (\eg~\cite{Fucci:2016}). \test ranges from 0 to $\infty$. 

We formulated the following parameterized null hypotheses,
\begin{description}
\item[HN1\textsubscript{X}.] There is no significant effect of Period with respect to  X  (\ie \qlty, \pro, or \test).
\item[HN2\textsubscript{X}.] There is no significant effect of  Technique with respect to X  (\ie \qlty, \pro, or \test).
\end{description}
The alternative hypotheses are two-tailed---\ie we did not consider the direction of the effect for either independent variable. 
HN1\textsubscript{X} was defined to investigate RQ1, while  we defined HN2\textsubscript{X} to investigate~RQ2.  


\subsection{Study Design}\label{sec:design}
The design of the cohort study is depicted in Table~\ref{tab:design}. The participants were randomly split into two groups---G1 and G2---having 15 participants each. Whatever the group was, the participants  were assigned to each treatment (\ie \tdd or \yw) twice. In particular, both groups were assigned to: \yw in the first period (\ie P1), \tdd in the second period (\ie P2),   \yw in third period (\ie P3), and  \tdd in the last period (\ie P4). Therefore, the design of our study can be classified as a \textit{repeated measures}, \textit{within-subjects} design. 
In each period, the participants in G1 and G2 dealt with different experimental objects. 
For instance, in P1, the participants in G1 dealt with \bsk, while those in G2  with \mra. 
At the end of the study, every participant had tackled each experimental object only~once. 

\begin{table*}[t]
\centering
\caption{Some descriptive statistics for each dependent variable grouped by Period and by Technique.}
\label{tab:stats}
\begin{tabular}{llcccccc} \toprule
\multirow{2}{*}{\textbf{Variable}}   & \multirow{2}{*}{\textbf{Statistic}} &\multicolumn{4}{c}{\textbf{Period (Technique)}} & \multicolumn{2}{c}{\textbf{Technique}} \\ 
   & &\textbf{ P1 (\yw)} & \textbf{P2 (\tdd)} & \textbf{P3 (\yw)} & \textbf{P4 (\tdd)} & \textbf{\yw} & \textbf{\tdd} \\ \midrule
\multirow{3}{*}{\qlty} &  mean &  59.3989 & 63.1002 & 63.0505 & 58.535  &  61.2247 & 60.8176 \\
&  median     &    76.7614 & 69.7251 & 71.2867 & 74.7614  &   72.9702 &  71.9962 \\
&  SD            & 37.8509 & 31.989 & 30.7322 & 34.5895 & 34.232 & 33.1112 \\ \midrule

\multirow{3}{*}{\pro} &  mean         & 34.1145 & 32.4793 & 30.991 & 37.9692 & 32.5527 & 35.2243 \\
 &  median         & 27.5281 & 29.0698 & 27.907 & 42.8571 & 27.907 & 34.8837 \\
&  SD & 32.182 & 29.039 & 28.9798 & 29.194 & 30.403 & 29.0012 \\ \midrule

\multirow{3}{*}{\test} &  mean         & 4.9333  &  7.8333  &  7.9333   & 10.1 & 6.4333                               & 8.9667 \\
&  median & 4 & 6.5 & 5 & 8.5 &  5 & 7 \\
&  SD & 4.0508 & 5.5216 & 7.5198 & 7.2462 & 6.1764 & 6.4885 \\ \bottomrule

\end{tabular}
\end{table*}

\begin{figure*}[t]
	\begin{subfigure}[t]{\columnwidth}
	\centering
  	\includegraphics[width=\columnwidth]{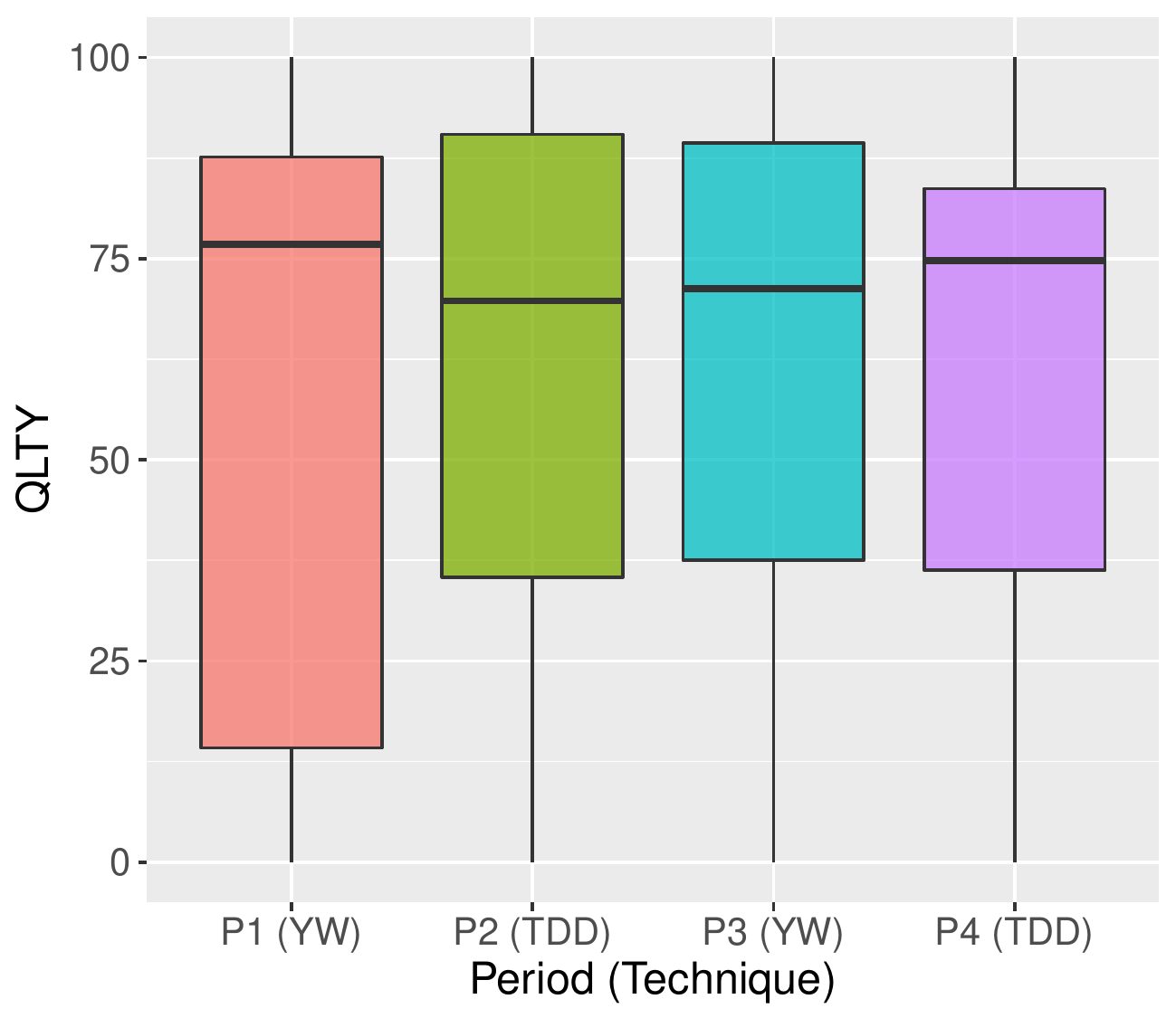}
    \caption{}
  	\end{subfigure}
    \begin{subfigure}[t]{\columnwidth}
	\centering
  	\includegraphics[width=\columnwidth]{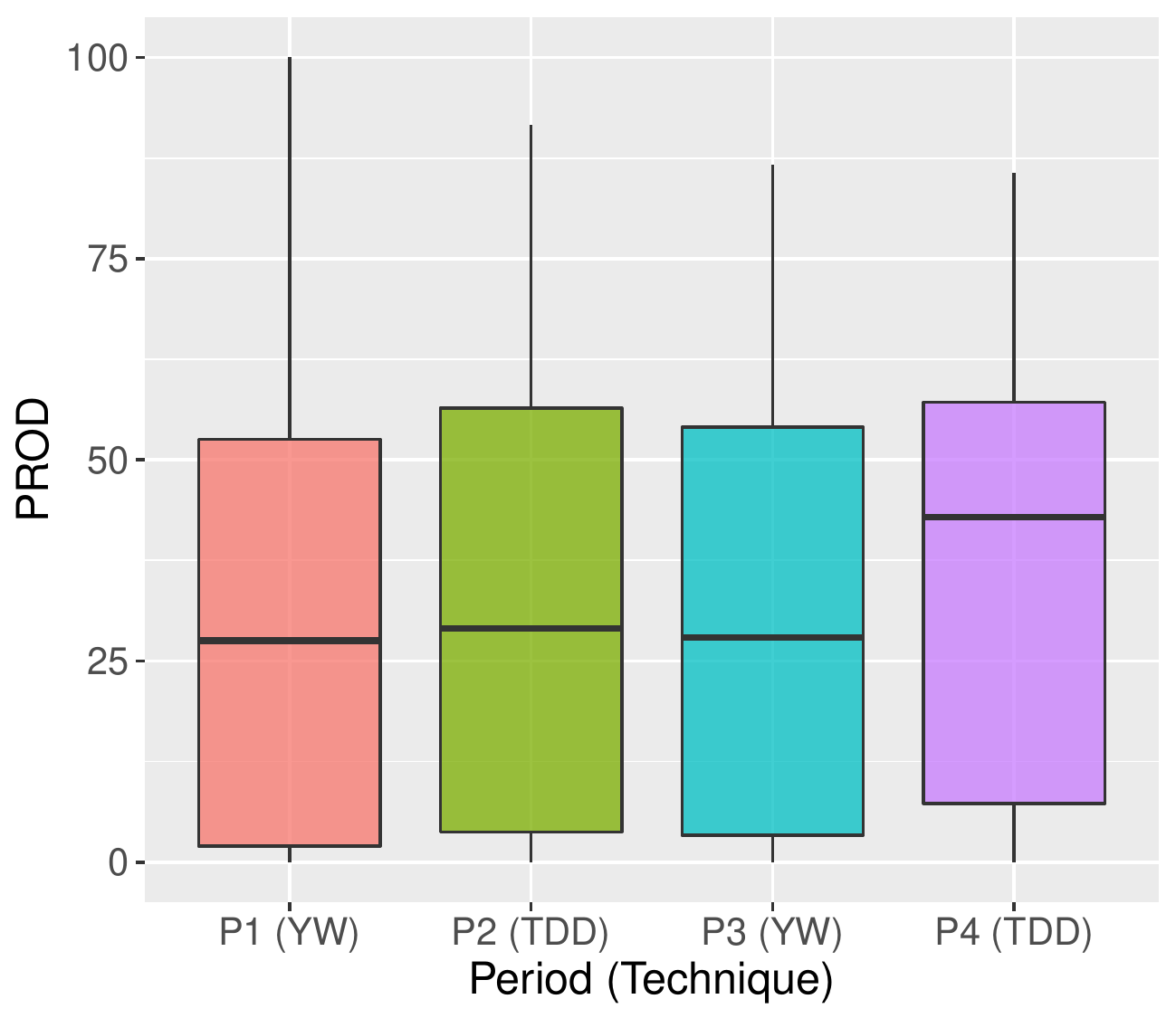}
    \caption{}
  	\end{subfigure}
    \begin{subfigure}[t]{\columnwidth}
	\centering
  	\includegraphics[width=\columnwidth]{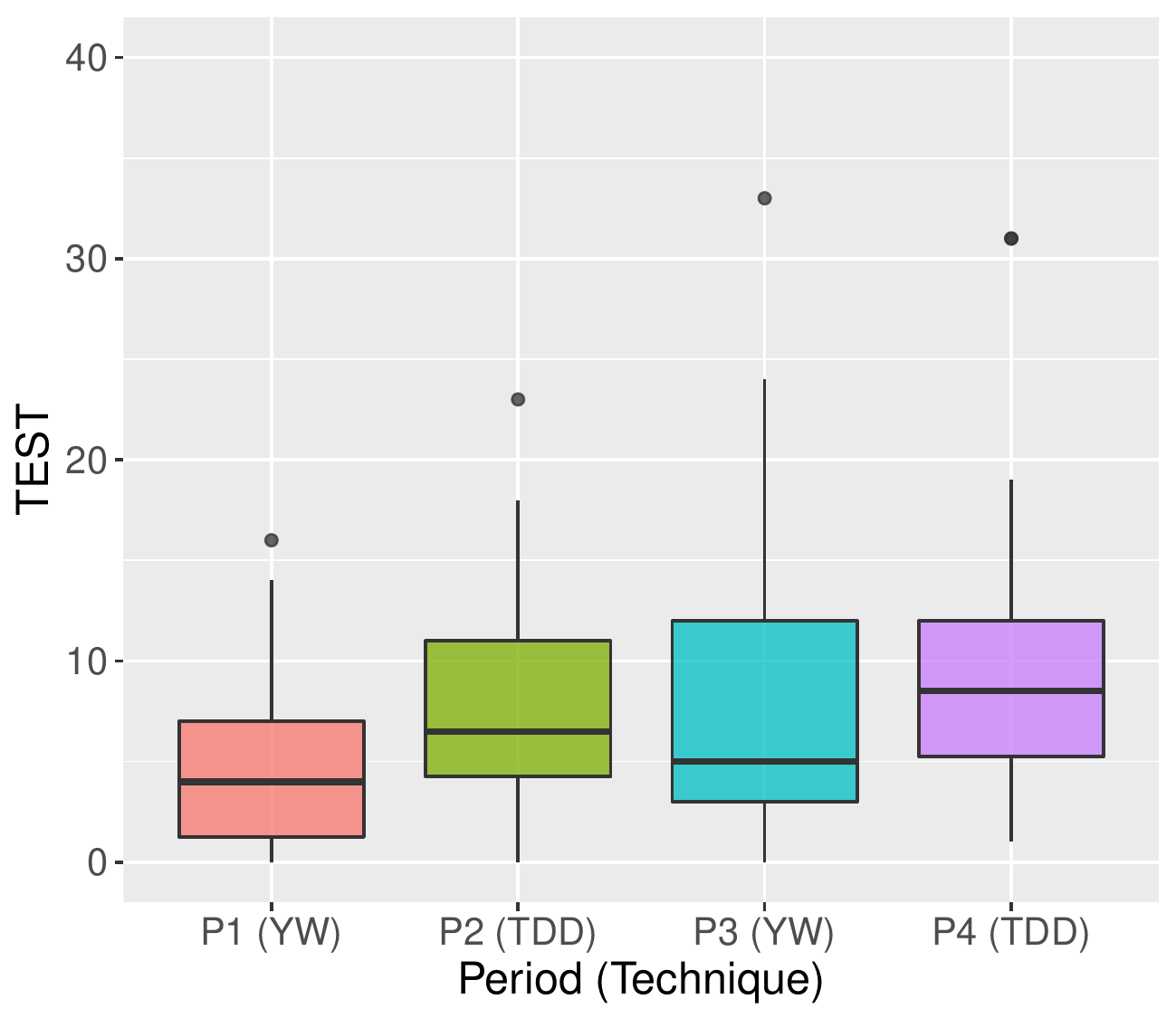}
    \caption{}
  	\end{subfigure}
  \caption{Boxplots for \qlty (a), \pro (b), and \test (c) for each period.}
  \label{fig:boxplots}
\end{figure*}

\subsection{Procedure}
Before our study took place, we collected some demographic information on the participants. To this end, the participants filled out an on-line pre-questionnaire (created by means of Google Forms). 

The Integration and Testing course---in which the cohort study was conducted---started in October, 2016. The first application of the \yw treatment (\ie P1) took place on November 16th, 2016 (see Table~\ref{tab:design}). Between the beginning of the course and P1, the participants had never dealt with \tdd, while they knew unit testing, iterative test-last development, and big-bang testing. On these techniques, the participants had taken part in two training sessions and  carried out  some homework. \tdd was introduced to the participants between P1 and P2. They also had taken part in three training sessions on \tdd and completed some homework by using this development practice. Given the previous considerations, we can exclude that the knowledge of \tdd affected in anyway the \yw treatment in P1. The \yw treatment was applied again on May 3rd, 2017 (in P3), while \tdd was applied the day after in P4. From P2 to P3 five months passed. Since the participants knew \tdd in P3, we cannot exclude that the knowledge of \tdd would have affected the treatment \yw in P3 in someway. That is, if the \tdd retainment had affected the application of \yw or not. On the other hand, we assessed the retainment of \tdd by asking the participants to use \tdd (once again) in P4. 

The execution of the study tasks took place under controlled conditions in a laboratory at the University of Bari. 
In each period, the participants in G1 and G2 were randomly assigned to the laboratory PCs. We alternated participants in G1 in G2 to avoid that participants of the same group assigned to the same experimental object were close to each other. This setup limited interactions among the participants. In addition, we  monitored them during the execution of tasks.  

All the PCs in the laboratory were equipped with the same hardware and software. Furthermore, they were set up with all the experiment materials necessary for carrying out the task, \ie the template project (of Eclipse) corresponding to the assigned experimental object. Each subject provided a solution for the assigned task by using the template project. The participants implemented the tasks in Java and used JUnit as testing framework. 
At the end of each task, the participants uploaded their solution through GitHub and then filled out a post-questionnaire. This questionnaire collected feedback on how the participant perceived the execution of each task.

\subsection{Analysis Procedure}\label{sec:analysis}
The gathered experimental data were analyzed according to the following procedure:
\begin{enumerate}
\item \textbf{Descriptive Statistics.} We computed descriptive statistics, \ie mean, median and standard deviation (SD), to summarize the distributions of the dependent variable values. We also used boxplots to graphically summarize these distributions. 

\item \textbf{Inferential Statistics.} We used Linear Mixed Model (LMM) analysis methods to test the defined null hypotheses. LMM is a popular method for analyzing data from longitudinal studies~\cite{Verbeke:2010}. For each dependent variable, we built an LMM that included the following terms: Period, Group, and Period:Group (\ie the interaction between Period and Group) as fixed effects, while the participant represents the random effect (this is customary in SE experiments~\cite{Vegas:2016}). It is worth noting that the periods P1 and P3 correspond to the application of the \yw treatment, while \tdd was applied in the periods P2 and P4. This means that, if the LMM does not indicate a statistically significant effect of Period, then there is no statistically significant effect of Technique. To build LMMs, we considered Group because, based on the study design, it also represents the sequence (\ie the order in which the treatments are applied in combination with the experimental objects). In repeated measures designs the effect of sequence on the dependent variables must be analyzed~\cite{Vegas:2016}. 

LMM analysis methods have two assumptions: \textit{(i)} the residuals of LMM have to be normally distributed and \textit{(ii)} their mean has to be equal to zero~\cite{Vegas:2016}. If these two assumptions are not verified, transforming the  data  of the dependent variable is an option (\eg by using log or square-root transformation)~\cite{Vegas:2016}. To check if the residuals were normally distributed, we used the Shapiro-Wilk test (Shapiro test, from here onwards)~\cite{Shapiro:1965}. As it is customary with tests of statistical significance, we accepted a probability of 5\% of committing Type-I error (\ie $\alpha = 0.05$).
\end{enumerate}

\section{Results}\label{sec:results}
In this section, we first report the results from the descriptive statistics followed by those pertaining the inferential statistics. 

\subsection{Descriptive Statistics}\label{sec:descriptive}
In Table~\ref{tab:stats}, we report the values of mean, median, and SD for each depended variable. These values are grouped by both Period and Technique. We also show the boxplots for the dependent variables in Figure~\ref{fig:boxplots}.  

\noindent
\textbf{\qlty.} 
As shown in Table~\ref{tab:stats} and Figure~\ref{fig:boxplots}.a, there are no noticeable differences in the \qlty values between the periods.
In particular, by comparing the boxplots for P1 and P3---\ie same treatment (\yw) but different experimental objects---we can observe that these boxplots overlap and the median level in P1 is higher than that in P3 (see also the median values in Table~\ref{tab:stats}). 

Similarly, we can observe that the boxplots for P2 and P4---\ie same \tdd treatment but different experimental objects---overlap and the median level is higher in P4. Therefore, such slight differences in the \qlty values seem to be due to the experimental objects rather than the retainment of \tdd.
Namely, when the experimental objects are \bsk and \mra (\ie in P1 and P4), the median levels are higher.   


When comparing \tdd and \yw, the results in Table~\ref{tab:stats} do not suggest differences in \qlty values (\eg on average, \qlty is equal to 60.8176 for \tdd, while it is equal to 61.2247 for \yw). This outcome is confirmed when we compare P4 (\tdd) with P1 (\yw) and P2 (\tdd) with P3 (\yw)---same experimental objects. For instance, the participants in P4 and P1 achieved, on average, similar values for \qlty (58.535 vs. 59.3989) although, when dealing with the same experimental objects, they applied either \tdd or \yw. The comparison between P2 and P3 lead to a similar observation.  

\noindent
\textbf{\pro.} 
The boxplots in Figure~\ref{fig:boxplots}.b do not indicate noticeable differences in the \pro values among the periods. Indeed, when passing from P2 to P4, we can observe that the boxplots overlap, but the median level for P4 is higher than for P2. In particular, the medians of  the \pro values are equal to 42.8571 and 29.0698 for P4 and P2, respectively. This improvement in the \pro values might be due to the \tdd retainment. As for the comparison between P1 and P3,  the boxplots are very similar to each other. Therefore, it seems that the knowledge the participants had of \tdd (\ie its retainment) in P3, with respect to P1, did not affect \qlty.

The results in Table~\ref{tab:stats} seem to suggest that there is a slight difference in the \pro values between \tdd and \yw in favor of \tdd (\eg  \pro for \tdd is equal to 35.2243 on average, while it is equal to 32.5527 for \yw ). By comparing pairs of periods in which the same experimental objects are used, we can observe that the \pro values in P4 (\tdd) are better than those in P1 (\yw). Namely, it seems that the participants who applied \tdd on \bsk and \mra achieved \pro values better than the participants who applied \yw. This trend is not observed when comparing P2 (\tdd) and P3 (\yw). For instance, the boxplot for P2 is very similar to that for P3, so suggesting that there is no difference with respect to the dependent variable \pro.

\noindent
\textbf{\test.} By looking at the boxplots in Figure~\ref{fig:boxplots}.c, we can observe differences in the \test values among the periods. In particular, if we compare the \yw treatments in P1 and P3, it appears that the boxplot for P3 is higher than that for P1. The \test values for P3 are also better on average (7.9333 for P3, 4.9333 for P1). This difference might suggest a positive effect of the \tdd retainment when participants had to apply \yw in P3. On the other hand, the boxplots for \tdd in P2 and P4 suggest a less pronounced difference in \test values. The boxplots for P2 and P4 overlap, although, as for P4, the boxplot is shorter and the median value is higher (8.5 in P4 vs. 6.5 in~P2).   

The comparison between \tdd and \yw seems to suggest that the \test values for \tdd are higher than those for \yw.  For instance, the mean values are equal to 8.9667 and 6.4333 for \tdd and \yw, respectively. By considering only P4 and P1, we can observe a clear improvement in the \test values in P4 (see the boxplots).
Namely, the participants who applied \tdd in P4 seem to achieve values for \test higher than those who applied \yw in P1 on the same experimental objects (\eg the mean values are 10.1 in P4, while 4.9333 in P1). Interestingly, the comparison between  P2 (\tdd) and P3 (\yw) does not confirm the trend previously observed. Namely, it seems that the distributions of the \test values for P2 (\tdd) and P3 (\yw) are quite similar (see both boxplots and descriptive statistics), despite the application of either \tdd (in P2) or \yw (in P3) on the same experimental objects. This outcome can indicate that the \tdd retainment influenced the participants who applied \yw in P3.

\begin{table}[t]
\centering
\caption{Results (\ie p-values) from the LMM analysis methods for the dependent variables \qlty, \pro, and \test.}
\label{tab:res}
\begin{threeparttable}
\begin{tabular}{lccc} \toprule
  \textbf{Variable} & \textbf{Period} & \textbf{Group}  & \textbf{Period:Group} \\ \midrule
\qlty & 0.8837 & 0.6108 & $<$0.0001\tnote{$\star$} \\
\pro  & 0.7973 & 0.8225 & $<$0.0001\tnote{$\star$} \\
\test & 0.0002\tnote{$\star$}  & 0.0617 & 0.4632 \\ \bottomrule       
\end{tabular}
\begin{tablenotes}
\item[$\star$] \footnotesize Statistically significant effect.
\end{tablenotes}
\end{threeparttable}
\end{table}

\subsection{Inferential Statistics}
The results (\ie p-values) from the LMM analysis methods are reported in Table~\ref{tab:res}. When a p-values is less than $\alpha$, we highlighted it with the $\star$ symbol.

\noindent
\textbf{\qlty.} The assumptions of the LMM analysis method for \qlty were both verified, so we did not perform any data transformation.  As shown in Table~\ref{tab:res}, the LMM analysis method does not allow us to reject HN1\textsubscript{\qlty}, the p-value for Period is 0.8837, namely the effect of Period is not statistically significant. 
This means that either there is no deterioration nor improvement in the observed time period (\ie no effect of time period) with respect to \qlty, or that the test does not have enough statistical power to show differences, would they exists.
The built LMM also suggests that the effect of Group is not statistically significant, while the interaction between Group and Period is. This interaction is due to the effect of the experimental objects (\eg whatever the treatment is, the distributions for \bsk are higher than those for \gol). 

Since LMM analysis method for \qlty does not indicate an effect of Period, the effect of Treatment is not statistically significant either. Therefore, we cannot reject HN2\textsubscript{\qlty}. 

\noindent
\textbf{\pro.} The LMM analysis method for \pro needed data transformation since the method assumptions were not satisfied. In particular, we applied a square-root transformation to meet these assumptions. The results in Table~\ref{tab:res} show that the effect of Period is not statistically significant. Therefore, we can neither reject HN1\textsubscript{\pro} nor HN2\textsubscript{\pro}, indicating that the participants may retain \tdd with respect to \pro. Moreover, applying either \tdd or \yw seems to not affect the \pro values.   
The LMM also includes a significant effect, namely Group:Period. Again, this significant interaction is due to the effect of the experimental objects.

\noindent
\textbf{\test.} To apply the LMM analysis method for \test, we had to transform the data of the dependent variable. In particular, we performed a log transformation so that the assumptions were verified.
The LMM analysis suggests that the effect of Period is statistically significant (the p-value is equal to 0.0002). Therefore, we can reject HN1\textsubscript{\test}. There is evidence that \textit{a significant effect of Period on the number of tests the participants wrote} exists. 
According to the boxplots in Figure~\ref{fig:boxplots}.c, the significant difference in Period is not due to a deterioration of \test values for \tdd over time---the worst distribution can be observed in P1---therefore, we can conclude that the ability of writing unit tests is retained by developers using \tdd. 

Since we found a significant effect of Period and in accordance with the results from the descriptive statistics (\ie there is a clear difference in favor of \tdd in P4 with respect to \yw in P1 on the same experimental objects), we reject HN2\textsubscript{\test}. Therefore, we can conclude that \textit{there is a significant effect of Technique on the number of tests the participants wrote}.

\section{Discussion}\label{sec:discussion}
In this section, we discuss the results obtained according to the RQs and present possible practical implications from our research. Finally, we delineate threats that could have affected the validity of our study. 

\subsection{Answers to Research Questions}
The data analysis gives some indication that developers retain \tdd. In particular, we observed neither deteriorations in the external quality of the solutions developed by the participants nor in their productivity. Moreover, it seems that, with time, there is an improvement in the number of tests written when using \tdd. 

Our results do not suggest differences between \tdd and \yw with respect to the quality of the implemented solutions, as well as the developer's productivity. However, who practices \tdd tends to write more tests. 

\subsection{Implications}
The participants retention of \tdd is particularly noticeable in the amount of unit tests written.
This is in line with the findings of a cross-sectional study by Erdogmus \etal~\cite{Erdogmus:2005} which pointed out that the number of tests correlates with the ability of novice developers to follow \tdd.
Our study extends that notion to a longitudinal perspective---\tdd helps retain \test over a period of five months.
The motivations for such effect are to be considered for further studies. However, we believe that \tdd raises the participants awareness about the importance of writing several (fine-grained) unit tests. 
Nevertheless, this does not translate in improved \qlty, nor \pro.
The latter result contrasts the ones of Latorre~\cite{Lat14} which saw, over a period of a month of \textit{constant} observation, a steady and significant improvement in performance measure similar to our \qlty.\footnote{In Latorre~\cite{Lat14}, all the subjects completed the task---\eg achieved \pro of 100\%.}
We conjecture that this can be the case due to the better experience of Latorre's study participants (\ie professional developers), furthering the thesis that \tdd alone is not a silver bullet but pre-existing skills play a crucial role~\cite{FTJ15,CSP11} 

Based on our findings, software companies that value unit testing (\eg for creating a regression for continuous integration) should encourage the use of \tdd as developers are likely to produce more tests when using such technique.
We showed that a small initial investment in training results in retainment of this particular feature on the long term.
Likewise, computer science educators should include \tdd early in their curricula to install a long-term unit-testing mentality in the students.
Finally, ``Experience with \tdd'' is a characteristic that researchers should foster when building a sample for studies requiring (novice) participants familiar with unit testing.
Likewise, when designing experiments where unit testing is desirable, researchers can avoid or limit time spent on training as, at least in our time-frame, such skill is retained by (novice) participants who already have minimal \tdd experience.  

Our results did not show any improvement of \tdd over \yw, contributing to the null results in \tdd research~\cite{Fucci:2016,fucci2013replicated,RM13}.
However, differently from previous attempts, we showed that no effects are observable also when the same subjects are tested again several months later, under similar conditions.
Time did not drastically decrement the novices performance when \tdd was applied, hinting at the fact that they soon regain familiarity with technique similarly to what the study of Latorre~\cite{Lat14} reports for the junior developers in the sample. 
Although carrying out cohort longitudinal studies---in particular, with several observations over a long time span---is difficult in SE (\eg controlling for maturation or learning effects), we put forward the idea that we might not be looking long enough (rather than hard enough) for the claimed effects of \tdd to become apparent.  
As a starting point towards this direction,
we recommend longitudinal studies in academia, which allow to follow the ``career'' of students over several years and can achieve a good amount of control (\eg based on grades, attendance)  

\subsection{Threats to Validity}
We discuss the threats that could have affected the validity of the obtained results according to the guidelines by Wohlin\etal~\cite{Wohlin:2012}. 
Accordingly, we ranked our threats from the most sensible for the goal of this study to the least one. 
In particular, being this the first test for a theory of TDD retainment, we prefer to limit threats to internal validity (\ie make sure that the cause-effect relationships are correctly identified), rather than being in favor of generalization.
 
\subsubsection{Threats to Internal Validity} This kind of threat concerns internal factors of the study that could have affected the results. 
The effect of letting volunteers take part in the study may influence the results because volunteers are generally more motivated~\cite{Wohlin:2012} (\ie \textit{selection threat}). 
To prevent participants exchanging information during the tasks (\ie \textit{threat of diffusion or treatments imitations}), at least two researchers monitored them.
We also prevented the diffusion of experimental materials by gathering it at the end of each task. 
A threat of \textit{resentful demoralization} might exist. For instance, a participant, who was given a less  desirable treatment or task, might not perform as good as they generally would. 
This threat to validity might have equally affected both \tdd and \yw.
Finally, control over subject \textit{maturation} was checked by making sure that the students attended the same courses between the first observation and the last one.    
\subsubsection{Threats to Construct Validity} These threats concern the relationship between theory and observation. The investigated constructs were quantified by means of one dependent variable each. This might affect the results (\ie \textit{threat of mono-method bias}). However, we used well-known and widely adopted dependent variables in \tdd experiments (\eg \cite{Fucci:2016}). Although the participants were not informed about the goals of the study, they might guess them and change their behavior accordingly (\ie \textit{threat of hypotheses guessing}). 
To mitigate an \textit{evaluation apprehension threat}, we told the participants that they would not be evaluated on the basis of their performances in the study. 

\subsubsection{Threats to Conclusion Validity} This kind of threat concerns the relationship between the dependent and independent variables. To mitigate a \textit{threat of random heterogeneity of participants}, our sample included students with similar backgrounds---\ie students taking the same courses in the same university with similar development experience. In empirical studies like this one, a \textit{threat of reliability of treatment implementation} might also exist. For example, a participant might follow \tdd more strictly than another one. We did not control for such effect in this study, however, we explicitly reminded the participants to follow the treatment they were assigned to.
The treatments might have impacted other constructs which were not observed (\eg number of refactoring, code complexity). 
Nevertheless, we focused on the most salient dependent variables according to the literature.
Finally, our sample was limited due to difficulty of recruiting participants available for the period of the entire study.

\subsubsection{Threats to External Validity} These threats concern the generalizability of the results. The participants in our longitudinal study were students, thus generalizing the obtained results to the population of professional developers poses a \textit{threat of interaction of selection and treatment}. However, the use of students as participants also implies a number of advantages~\cite{Carver:2003}, such as participants with homogeneous background, the possibility to obtain preliminary evidence. 
In addition, the tasks to be performed did not require a high level of industrial experience, we believe that the use of students as participants could be considered appropriate, as suggested in the literature~\cite{Carver:2003,Host:2000}. 
The experimental objects might also affect the external validity of the results (\ie \textit{threat of interaction of setting and treatment}) as they are not representative of real-world settings.
On the other hand, simpler tasks, which can be completed in a single exercise session (approximately three hours), allow a better control over the participants. The latter was our preferred trade-off due to the theory-testing nature of this study.

\section{Conclusion}\label{sec:conclusion}
In this paper, we present a quantitative longitudinal cohort study to investigate: \textit{(i)} the \tdd effects on the external (\ie functional) quality of software products as well as the developers' productivity; and \textit{(ii)}~the retainment of \tdd over a period of five months. 
The results indicate that the use of \tdd has a statistically significant effect neither on the external quality of software products nor on the developers' productivity. 
However, we observed that participants using \tdd produced significantly more tests than those applying a non-\tdd development process, and that this capacity was retained over time.
On the basis of these findings, we speculate that software companies can be encouraged to adopt \tdd because \textit{(i)} it compels developers to write more unit tests, which can be leveraged for localizing faults through regression and \textit{(ii)} it requires minimal initial effort to retain its effect. 

There are several future directions for the research presented in this paper.
First, we will replicate this study in academic context, with the same cohort but on a longer time-span with several observations.
Second, we will replicate the same design presented here with professional developers---\eg in the form of workshops about Agile software development. 
Third, we will devise qualitative longitudinal studies to triangulate statistical results.

\bibliographystyle{ACM-Reference-Format}
\bibliography{sample-bibliography}

\end{document}